\documentclass[aps,prl,twocolumn,showpacs,a4paper,superscriptaddress]{revtex4}
\usepackage{graphicx}
\usepackage{amsmath}
\begin{document}
\title{Mechanical mode dependence of bolometric back-action in an AFM microlever}

\author{G. Jourdan}
\affiliation{Institut N\'eel, CNRS-UJF, BP 166 38042, Grenoble Cedex 9, France\\}
\affiliation{Universit\'e Joseph Fourier, BP 53 38041, Grenoble Cedex 9, France\\}
\affiliation{Laboratoire Kastler Brossel, CNRS-ENS-UPMC, 4 Place Jussieu, 75252 Cedex 05, France\\}

\author{F. Comin}
\affiliation{ESRF, 6 rue Jules Horowitz, BP220, 38043 Grenoble Cedex, France\\}

\author{J. Chevrier}
\affiliation{Institut N\'eel, CNRS-UJF, BP 166 38042, Grenoble Cedex 9, France\\}
\affiliation{Universit\'e Joseph Fourier, BP 53 38041, Grenoble Cedex 9, France\\}
\affiliation{ESRF, 6 rue Jules Horowitz, BP220, 38043 Grenoble Cedex, France\\}

\date{\today}

\begin{abstract}

Two back action (BA) processes generated by an optical cavity based detection
device can deeply transform the dynamical behavior of an AFM microlever: the
photothermal force or the radiation pressure. Whereas noise damping or
amplifying depends on optical cavity response for radiation
pressure BA, we present experimental results carried out under vacuum
and at room temperature on the photothermal BA process which appears
to be more complex. We show for the first time that it can simultaneously
 act on two vibration modes in opposite direction: noise on one mode is
amplified whereas it is damped on another mode.
Basic modeling of photothermal BA shows that dynamical effect on mechanical mode is laser spot position dependent with respect to mode shape. This analysis accounts for opposite behaviors of different modes as observed.

\end{abstract}


\maketitle

Cooling down the main degree of freedom of a micro mechanical resonator has been the recent focus of numerous studies aimed at reaching its quantum ground state \cite{Cohadon,Karrai,Arcizet_N2006,Gigan,Kleckner,Schliesser,Naik,Corbitt,Favero,Poggio,Thompson}. Besides such an experimental challenge, the prospect of building entangled quantum state between macroscopic object and photon, spin or electron opens new ways towards quantum information and to some extent towards classical to quantum behavior boundary study \cite{Schwab,Marshall}. Highly sensitive measurement of small displacement is limited by quantum BA \cite{Arcizet_PRA2006}. For instance it sets the standard quantum limit of interferometer developed for gravitational wave detectors through Heisenberg relationship, that links phase measurement and radiation pressure. Some research aimed at getting around such a major limitation are considering detuned cavity to reach the ultimate quantum limit only related to mechanical dissipation of mirrors \cite{Arcizet_PRA2006,Braginsky}. 

Here we present self cooling of an AFM lever by means of photothermal force: as quoted in \cite{Gigan}, such force can participate to cooling of an oscillator even at high frequency (larger than 100 kHz). Moreover in \cite{Metzger} possibility of photothermal BA to cool down a mechanical oscillator to its quantum ground state is discussed.
Here we show opposite back-action effects on various mechanical modes: to understand such a behavior, optomechanical coupling process needs a comprehensive description. Dynamical radiation pressure effect is mainly determined by optical cavity response \cite{Wilson} and mechanical modes parameters.
In case of photothermal process, optomechanical coupling is on top of that monitored by material structuring of oscillator and temperature field distribution related to laser beam position with respect to mode shape.
In the here presented experiment, mode 1 is warming up, while mode 0 is cooling down and vice versa. Such behavior could be prejudicial for mode cooling efficiency, since modes are actually weakly coupled to each other through BA process \cite{Arcizet_N2006}. Opposite effects generated by competition between photothermal and radiation BA can also occur on a mechanical mode, since it is shown that the first one is position dependent on the system.\\
\begin{figure}[b]
	\centering
		\includegraphics[width=0.49\textwidth]{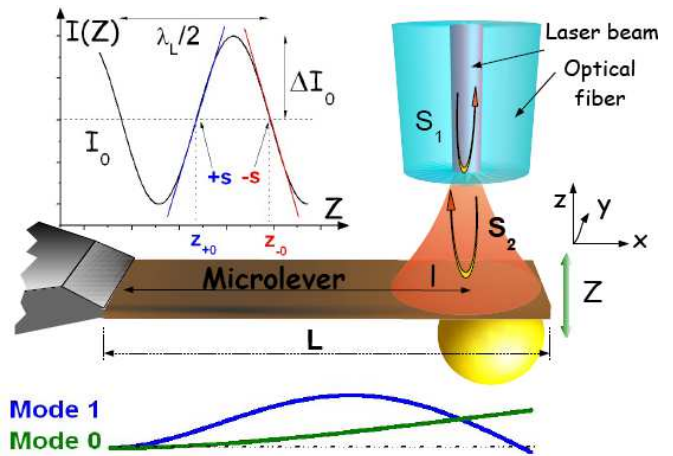}
	\caption{The optical fiber based interferometer is sensitive to oscillator motion. A microsphere is glued on the lever, thus placing mode 1 node almost at the end of the structure. The back of the lever and the optical fiber end are forming a poor finesse cavity: as indicated in the inset, intracavity intensity is cavity length dependent with period $\lambda/2$, where $\lambda_L = 670 $ nm is the laser wavelength. Modulation $\Delta I_0/I_0$
	 is nevertheless expected to be very weak.}
	\label{fig:figure1}
\end{figure}
In our experiment, the mechanical resonator consists of an AFM 300 nm thick gold coated microlever \footnote{MikroMasch CSC17: $L \times w \times t = 450 \times 50 \times 2$ $ \mu m^3$} with a 40 $\mu$m radius sphere glued at its end, dedicated to Casimir force study \cite{Jourdan}. The first two resonance frequencies amount to $f_0=3943.5$ Hz and $f_1=38443.5$ Hz. Under vacuum ($P \approx 10^{-6}$ Torr) and at room temperature, dissipation rates related to thermal bath coupling are respectively $\gamma_0=12.3$ rad.s$^{-1}$ and $\gamma_1=95$ rad.s$^{-1}$. 
An optical fiber based interferometer is implemented in order to measure the oscillator motion (Fig.~\ref{fig:figure1}). A laser beam led by an optical fiber is reflected off the microlever and then coupled back to the same optical fiber to generate a two waves interference signal at the photodiode level. However intracavity intensity modulation is nevertheless expected even if it is very weak. this is essentially due to laser losses associated to successive reflections that increasingly arise as a result of beam section enlargement. Therefore inside the cavity defined by the microlever and the optical fiber end, intensity is affected by mirror distance. Spatial shape of intensity distribution can therefore be rather complex as suggested by cavity detuning study: it exhibits a $\lambda_L/2$ period along z axes. 
When collecting the optical motion signal, the cavity length is classically set at its maximum motion sensitiveness ($z_{\pm 0}$) and then stabilized against mechanical drift by means of a piezo transducer associated to the fiber, fed back by the low frequency part ($<$ 1 Hz) of the motion signal. It ensures position sensitiveness and mechanical response of oscillator, as described below, to remain the same.\\
Depending on working position inside optical cavity ($z_{\pm 0}$ in Fig.~\ref{fig:figure1}), lever mechanical response exhibits two opposite behaviors. As indicated in Fig.~\ref{fig:figure2}, dissipation rates and resonance frequencies of mode 0 and 1 depend linearly on laser beam intensity: the stronger the field, the larger the discrepancy with respect to the undisturbed lever response. Changes in mechanical parameters are caused by thermal force that is sensitive to intracavity intensity variation when the lever is moving: this BA process has been reported for the first time in \cite{Karrai} and then observed in \cite{Gigan}. It should be noted that lever damping by photothermal force had been previously observed in \cite{Mertz} by means of external feedback loop that modulated intensity of a laser hitting a lever. 
Here in Fig.~\ref{fig:figure2} (a and b), for position $z_{+0}$, we observe that mode 0 dissipation rate is going up while mode 1 dissipation rate is decreasing. As a result, in Fig.~\ref{fig:figure3}, Brownian motion of mode 0 is damped while simultaneously it is enhanced for mode 1. Conversely, for position $z_{-0}$, opposite observation can be made. So far, opposite effects on various modes have not been shown to our knowledge. It actually derives from a more comprehensive context: unlike pressure radiation, thermal force is generated through a process that takes place on the whole mechanical structure. Local and non local force generating account for this major distinctness.
Radiation pressure is here not considered. Because of the low cavity finesse, its dynamical back-action can not account for dissipation rate change. However optical spring can participate to frequency shift observed although its relative contribution on mode 0 is expected to be smaller than 1 $\%$. Intensity modulation is indeed rather low inside the cavity.\\
\begin{figure}[t]
	\centering
		\includegraphics[width=0.49\textwidth]{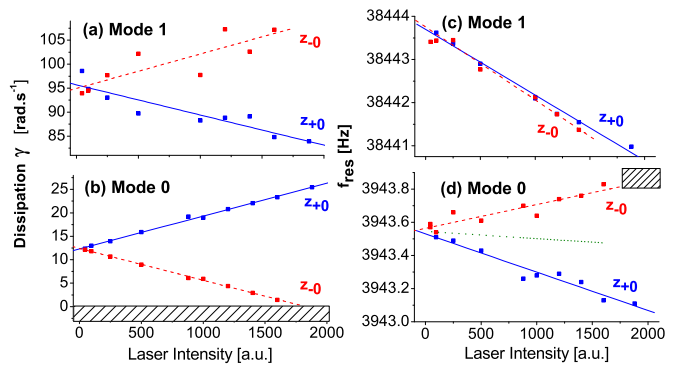}
	\caption{In graph (a) and (b), dissipation rates for mode 0 and 1 are plotted against Laser intensity in arbitrary unit, at each cavity detuning $z_{+0}$ and $z_{-0}$. At 2000 a.u. optical power is in the order of 500 $\mu$W. Graph (c) and (d) display resonance frequency shift generated by thermal force and lever heating: for mode 1, latest process appears to be the dominant one, since frequency shift is decreasing in both case whatever the cavity detuning. Graph (d) reveals change of resonant frequency produced by heating through the dotted line: as a result, it accounts for slope difference between the two branchs $z_{+0}$ and $z_{-0}$. Hatched area is related to instability behavior we observed on mode 0 for $z_{-0}$, when making $\gamma_0$ negative. Mechanical parameters for mode 0 and 1 are acquired simultaneously through thermal mechanical noise analysis: Brownian motion peaks are fitted with Lorentzian shape curve, whose parameters are $\omega_i$, $\gamma_i$, curve area $<z^{2}_{i}>$ and pedestal $y_i$.  }
	\label{fig:figure2}
\end{figure}
Lever thermal activation arises as a result of thermoelastic expansion of solid lattice when changing temperature. Laser beam used to probe microlever motion is partially absorbed with rate A. It induces local temperature increase that then takes place on the whole system through thermal energy diffusion. 
As thermal force is mainly defined by structure material and temperature field distribution $\Delta T(x,z)$, delay in temperature propagation may result in delay in force with respect to laser intensity change.
Complementary experiments performed in air on same microlever model consisted in irradiating the oscillator with a modulated intensity laser beam at frequency $\omega$.
Mechanical response analysis showed that thermal force exhibits a first order low pass behavior at least for the first two modes of the lever (slope $-20$ dB$/$dec):
\begin{equation}
 G(\omega)=\frac{F_{th}}{A \Delta I_L}=\frac{\beta}{1-j\omega \tau_c}
 \label{eq:Lowpass_force}
\end{equation}
Temperature distribution can be estimated within the heat equation framework. If one considers a homogeneous material for whole lever, temperature is qualitatively expected to decrease exponentially and to oscillate along the beam with length scale $\lambda_t = \sqrt{2 \gamma_t/ \omega c_t}$, where $\gamma_t$ and $c_t$ are thermal conductivity and heat capacity. Moreover, temperature profile should be proportional to $\lambda_t$, because energy flux input is here imposed. 
In fact propagation effect is negligible over length scale $\lambda_t$: it means that thermal force is not delayed if $\lambda_t(\omega) > L$, the lever length, \textit{ie} $\omega<2 \gamma_t / (L^2 c_t)=1/\tau_t$. For silicon or gold material \footnote{ Si: $\gamma_t=148$ W/m/K, $c_t=1.63 \; 10^6$ J/m$^3$/K } \footnote{ Au: $\gamma_t=317$ W/m/K, $c_t=2.47 \; 10^6$ J/m$^3$/K }, $\tau_t$ is evaluated at 1.1 ms and 0.8 ms, which is in good agreement with order of magnitude of the response time $\tau_c = 1.6$ ms associated to cut frequency $f_c \approx 100$ Hz we observed in the experiment above.\\
When oscillating at frequency $\omega$, the lever experiences power absorption proportional to its motion $Z$: $A \Delta I_L = \pm s A Z$ for cavity position $z_{\pm 0}$ (see inset in Fig.~\ref{fig:figure1}: $\pm s$ is intracavity intensity slope against cavity length). Therefore it generates a thermal force as described by Eq.~(\ref{eq:Lowpass_force}). Around resonance frequency $\omega_i$ of mode \mbox{$i=0,1$}:
\begin{equation}
F_{th}=\frac{\pm s A \beta}{1+(\omega_i \tau_c)^2}(Z- \tau_c \dot{Z})= \pm(\Delta k Z + \Delta \Gamma \dot{Z})
\label{eq:Thermal_force_position}
\end{equation}
It induces change in damping rate $\Delta \Gamma = m \Delta \gamma$ as well as in oscillator stiffness $\Delta k = 2 m \omega_{res} \Delta \omega_{res}$. Since slope $\pm s$ is expected to be proportional to laser intensity, Eq.~(\ref{eq:Thermal_force_position}) accounts for damping rate shift in Fig.~\ref{fig:figure2}: for both modes, $\Delta \gamma (z_{+0}) = -\Delta \gamma(z_{-0})$. However, Eq.~(\ref{eq:Thermal_force_position}) does not take fully into account resonance frequency shifts observed in Fig.~\ref{fig:figure2} for mode 0 and 1: lever temperature is increasing, thus causing  resonance frequency to drop, mainly because of Young modulus thermal sensitivity. For silicon cantilever, shift is expected to be $(\partial f_{res} / \partial T)/f_{res}=-5.2 \; 10^{-5} K^{-1}$ \cite{Cleland,Giessibl}.
In Fig.~\ref{fig:figure2}c and~\ref{fig:figure2}d, major feature of the mode 0 and 1 behaviors have completely different origin. For mode 1, independent of cavity state ($z_{\pm 0}$), the resonance frequency decreases: this is due to lever temperature. For mode 0, resonance frequency increases ($z_{-0}$) or decreases ($z_{+0}$) depending on cavity state. Mode 0 is first sensitive to self cooling effect. A detailed analysis in Fig.~\ref{fig:figure2}d (dotted green line) however shows a residual thermal effect.
 At maximum intensity, for mode 1, heating is estimated around 1 K, which is consistent in order of magnitude with intensity absorption around $30 \; \mu W$, given thermal parameters of the structure. By subtracting heating effect in data for mode 0, one can evaluate delay time $\tau_c = 0.1$ ms in Eq.~(\ref{eq:Thermal_force_position}), since $\Delta \gamma_0 = - 2 \omega_0 \tau_c \Delta \omega_0$ with $2 \omega_0 \tau_c = 5.5$. It appears to be in good agreement with previous estimation or evaluation ($\tau_t, \tau_c \approx 1$ ms), given simplicity of model Eq.~(\ref{eq:Lowpass_force}).\\
A simple model for thermal activation is now developed.
It is based on classical treatment of beam deflection that assumes inner parts of the system to be stress free except along the beam direction. Comprehensive description should include other thermal stress components that can bring certainly non negligible contributions. 
 Mechanical stress $\sigma$ is locally defined by deformation $\varepsilon$ and temperature $T+\Delta T$ within the thermoelastic Hookes law: $\sigma = E (\varepsilon - \alpha \Delta T)$ where $E$ is the Young's modulus and $\alpha$ the thermal expansion coefficient. Following the standard derivation procedure, the equation of motion of thermoelastic beam can be drawn \cite{Lifshitz}. Transverse vibration mode equation for $a_n$ results from the projection of beam deformation $Z(x,t) = \sum U_n(x) a_n(t)$ on mode shape $U_n(x)$:
\begin{equation}
m \ddot{a_n} + \Gamma_n \dot{a_n} + m \omega_{n}^{2} a_n= F_{th,n} = - \int_{0}^{L} E \frac{\partial^2 U_n}{\partial x^2} I_T dx
\label{eq:thermal_force_mode_n}
\end{equation}
$m$, $\omega_{n}^{2}$ and $\Gamma_n$ are effective mass, resonance frequency and damping rate associated to mode $n$. The force is generated all along the beam with length $L$ through thermal contribution to moment of inertia integrated over cross section $I_T = \int z \alpha(z) \Delta T(x,y,z) dy dz$. Here thermal bimorph effect is roughly taken into account through the $z$ dependence of $\alpha$: $I_T$ is made non zero for homogeneous temperature distribution over the cross section. For simplicity, Young's modulus is assumed to be the same over the whole section. In one material made microlever, thermal actuation of transverse mode is mainly explained by temperature gradient along $z$ axis. Flexural and longitudinal mode, excited through the same thermal process described above, are uncoupled for small motion of naked beam. We neglected the coupling generated by the out of beam deported mass of the microsphere at the end of lever, since resonances are expected to be away from each other. 
\begin{figure}[t]
	\centering
		\includegraphics[width=0.49\textwidth]{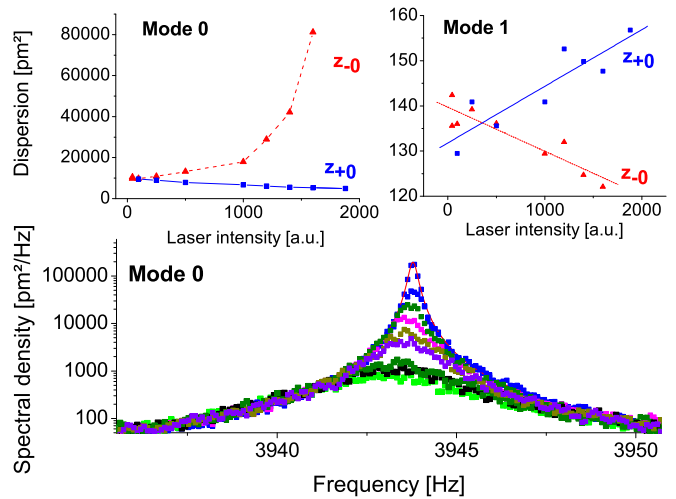}
	\caption{In graphs (a) and (b), Brownian motion of mode $i$, $i=0,1$, $<z^{2}_{i}>$ is plotted against laser intensity, which is proportional to damping rate $\gamma_i$. For each cavity position ($z_{\pm 0}$), mechanical noise suits relation \mbox{$T_{eff}=k <z^{2}_{i}>/k_B=\gamma_i/(\gamma_i+\Delta \gamma_i)$}. Graph (c) displays noise spectrum density around fundamental resonant frequency, when increasing laser intensity. According to cavity detuning, Brownian motion is damped ($z_{+0}$) or enhanced ($z_{-0}$).
	        }
	\label{fig:figure3}
\end{figure}
Temperature distribution can be described as $\Delta T(x,z,t) = D_l(x,z,t)\otimes A \Delta I(t)$, where $l$ denotes the laser ray position on the lever and $D_l(x,z,t)$ is the Green's function associated to heat equation. Power absorption $A \Delta I(t)$ is equal to  $\pm s A \sum U_{n'}(l) a_{n'}(t)$. Thermal force on mode $n$ is then evaluated through Eq.~(\ref{eq:thermal_force_mode_n}): it generates changes in mechanical parameters of oscillator $n$ and coupling between modes $n$ and $n'$.
\begin{equation}
F_{th,n}= \pm s A G_l^n(t) \otimes \left( U_n(l) a_n + \sum_{n \neq n'} U_{n'}(l) a_{n'} \right)
\label{eq:thermal_force_an}
\end{equation}
where $G_l^n(t)=-\int E \frac{\partial^2 U_n}{\partial x^2} z \alpha D_l(x,z,t) dx dy dz$. Not surprisingly we observe in second term of Eq.~(\ref{eq:thermal_force_an}) that self cooling inherently introduces mode coupling that has experimentally limited effects. Damping or enhancing of mode $n$ is directly related to sign of Eq.~(\ref{eq:thermal_force_an}) with respect to $a_n$, which is defined by $\pm s U_n(l) G_l(t)$. When crossing a vibration node, $U_n(l)$ changes sign, whereas $G_l(t)$ should not in most case, since temperature distribution should remain almost the same inside the lever. This clearly shows that BA effect on mechanical responses can be different simultaneously on modes 0 and 1. In our experiment, laser spot was located between the lever basis and node of mode 1, since associated noise was decreasing when the spot was shifted to the end of the lever. Because the sphere mass put the node almost at the extremity, we were unable to cross it and observe opposite BA effect. However such an observation on similar system has been reported in \cite{Ortlieb}.
Now effect of radiation pressure back-action on various modes at frequency $\omega_n$ is conveniently analysed using the intracavity intensity response 
$R(\omega_m)= \Delta I(\omega_m)/Z(\omega_m)$. Force density on the lever is indeed described by $F(x)= 2 \delta(x-l) \Delta I(t)/c $, thus exerting on mode $n$ the force $F_n$ (without mode coupling components on the right term):
\begin{equation}
F_n= \int F(x) U_n(x) dx \approx \frac{2 U^2_n(l) R(t)}{c} \otimes a_n
\label{eq:radiation_pressure_mode_n}
\end{equation}
  $c$ is light velocity. Such description is valid as long as laser spot size is smaller than mode $n$ deformation length $\lambda_n$ associated to $U_n$. $F_n$ proportional to $U^2_n$ can be zero but cannot change sign. Contrary to photothermal process, cavity response $R(\omega)$ solely determines sign of back-action: when crossing node, the latter should in particular remain the same.\\
BA generated by thermal force is not only laser spot position dependent: Eq.~(\ref{eq:thermal_force_an}) suggests also that damping rate variation depends on frequency $\omega$ through $G_l^n$. Two extreme cases can be considered: first, at low frequency, when $\lambda_t(\omega) >> L$, temperature field is almost homogeneous on the lever, thus producing force proportional to $\pm s U_n(l) \frac{\partial U_n}{\partial x}(L)$. When working between lever basis and node of mode 1, it shows that BA effects are opposite on modes 0 and 1. At higher frequency, ie when $\lambda_n>>\lambda_t(\omega)>>w$, temperature disturbance is concentrated around laser spot position $l$ with extension $\lambda_t$ along the beam, but should remains homogeneous across the beam section $w$. Thermal force is expected to be  proportional to $\pm s U_n(l) \frac{\partial^2 U_n}{\partial x^2} (l)$. As a result, sign of damping rate shift can be opposite to previous case.\\
As a conclusion, starting from the model here described, using a response function $G_l^n(\omega)$, we can emphasize that thermal force BA on mode $n$ is laser spot position dependent as well as frequency dependent. Such a behavior can raise major issue in the prospect of mode cooling. Thermal force may indeed introduce instability on a specific mode, when cooling down another one. As mentioned above, BA generates coupling between various modes. In case of radiation pressure process, in \cite{Arcizet_N2006} it is shown that cooling efficiency is better when taking into account mechanical noise background generated by other modes. In case of photothermal process, enhancement of mechanical noise background, produced by mode instability, could impose a limitation to oscillator cooling, beside heat absorption issue mentioned in \cite{Metzger}.\\
Temperatures achieved for the two modes in our setup are asking for comments. $T_{eff}=T/2$ is obtained for mode 0: temperature drop is limited by maximum intensity of the laser. BA process turns out to be much less efficient for mode 1, since temperature shift achieved is only $15\%$ with respect to room temperature. Photothermal effect emphasizes once more how much self cooling depends on details of opto mechanical coupling. More important than this quantitative difference is the central result presented in this paper: 
photothermal BA can act in opposite directions on various vibration modes.\\
Support for this work was provided by the European contract STRP 505634-1 X-Tip.

\end{document}